\documentclass[doublespacing]{elsart}

\usepackage{graphicx}
\usepackage{amssymb}

\begin{document}

\begin{frontmatter}


  % use the thanksref command within \title, \author or \address for
  % footnotes; use the corauthref command within \author for
  % corresponding author footnotes; use the ead command for the email
  % address, and the form \ead[url] for the home page:
  % \title{Title\thanksref{label1}} \thanks[label1]{}
  % \author{Name\corauthref{cor1}\thanksref{label2}} \ead{email
  %   address} \ead[url]{home page} \thanks[label2]{} \corauth[cor1]{}
  % \address{Address\thanksref{label3}} \thanks[label3]{}

  \title{Active control of sound inside a sphere via control of the
    acoustic pressure at the boundary surface}

  % \author[label1,label2]{} \address[label1]{} \address[label2]{}

  \author{N. Epain}, \ead{epain@lma.cnrs-mrs.fr} \author{E. Friot}
  \ead{friot@lma.cnrs-mrs.fr} \address{Laboratoire de M\'{e}canique et
    d'Acoustique\\31, chemin Joseph Aiguier\\13402 Marseille cedex 20
    France}

  \begin{abstract}
    % Text of abstract
    Here we investigate the practical feasibility of performing
    soundfield reproduction throughout a three-dimensional area by
    controlling the acoustic pressure measured at the boundary surface
    of the volume in question. The main aim is to obtain quantitative
    data showing what performances a practical implementation of this
    strategy is likely to yield. In particular, the influence of two
    main limitations is studied, namely the spatial aliasing and the
    resonance problems occurring at the eigenfrequencies associated
    with the internal Dirichlet problem. The strategy studied is first
    approached by performing numerical simulations, and then in
    experiments involving active noise cancellation inside a sphere in
    an anechoic environment. The results show that noise can be
    efficiently cancelled everywhere inside the sphere in a wide
    frequency range, in the case of both pure tones and broadband
    noise, including cases where the wavelength is similar to the
    diameter of the sphere. Excellent agreement was observed between
    the results of the simulations and the measurements. This method
    can be expected to yield similar performances when it is used to
    reproduce soundfields.
  \end{abstract}

  \begin{keyword}
    soundfield reproduction \sep active noise control \sep three
    dimensional \sep boundary pressure \sep internal Dirichlet problem

    \PACS 43.38.Md \sep 43.60.Tj \sep 43.50.Ki
  \end{keyword}
\end{frontmatter}

% main text
\section{Introduction}
\label{introduction}
During the last few decades, considerable attention has been paid to
soundfield reproduction, which raised many issues in acoustics and
signal processing. Several methods can be used to reproduce a given
soundfield, the most popular of which are the binaural techniques
\cite{Begault}, Ambisonics \cite{Gerzon}, and Wave Field Synthesis
\cite{Berkhout}. For a given application, the choice of method will
depend on the physical properties of the sound to be reproduced and on
the number of listeners.
\par
We focus here in particular on the reproduction of very low frequency
soundfields. One of the practical applications of reproducing
soundfields of this kind is the perceptual assessment of sonic boom
sounds. It is particularly difficult to choose a suitable method of
reproducing these soundfields because of their spectral properties:
most of the energy of which sonic boom sounds consist involves
frequencies below a few dozen~Hz, and the spectrum is maximum at only
a few~Hz \cite{Dancer,Coulouvrat}. Besides, the acoustic pressure
level of sonic booms is often as high as $120$~dB.
\par
More generally, reproducing low-frequency soundfields is of interest
for the study of hearing. Perceptual assessments such as those
focusing on the inconvenience associated with transport noise are
usually performed using headphones. This is the simplest and cheapest
method available to deliver a desired soundfield to the ears of a
listener. However, this method is not suitable for reproducing very
low frequencies, which are thought to be perceived not only by the
ears themselves, but by the entire body. The soundfields therefore
have to be extended spatially so that they surround at least the
listener's head and torso.
\par
Contrary to binaural techniques, Wave Field Synthesis (WFS) and
Ambisonics both give a suitably large reproduction zone surrounding
the listener. However, the main problem on which Ambisonics focuses is
not so much to accurately reproduce soundfields, but to give the
listener a realistic spatial feeling. The principle on which the
method is based involves several psychoacoustic hypotheses, which
raises problems in the case of infrasonic sounds since the hearing
mechanisms mobilised at these frequencies have not yet been completely
elucidated. Unlike Ambisonics, WFS focuses on reproducing the physical
properties of soundfields in a wide spatial area. Unfortunately, this
technique was originally unable to compensate for the reflective
properties of the sound reproduction room, whereas reproducing sonic
booms requires using a small room to reach the pressure levels
required at such low frequencies. Recent studies \cite{Corteel,Spors}
have shown that a preliminary equalization step can be used to correct
the errors induced in WFS by room reflections. However the procedures
proposed for this purpose are open-loop ones, which means that the
performances of the system can be affected by any change in the
physical properties of the reproduction room, such as temperature
changes.
\par
Several sound reproduction methods have been proposed since the
nineties, based on active noise control strategies. Cancelling a
primary noise is basically the same task as reproducing it, since a
perfect cancellation of the primary noise requires generating a
secondary noise which has exactly the same properties, apart from they
are in opposition of phase. The main advantage of methods of this kind
is that they can be used to compensate for the reflections generated
in the reproduction room by means of adaptive filters. Some of these
``active control'' methods can be compared with local active noise
control strategies
\cite{Kirkeby_1,Kirkeby_2,Nelson_jsv,Kahana,Bauck,Kirkeby_3,Choi}.
With these methods, either error microphones are placed directly in
the area where the soundfield is to be reproduced, or preliminary
measurements are carried out to design control filters. The presence
of microphones in the reproduction area is not advisable because these
conditions may be uncomfortable for the listener. In addition, in
order to reproduce a soundfield in a three dimensional area, a very
large number of sensors can be required. On the other hand, as with
WFS, the use of an off-line filter design prevents the performances of
the system from remaining constant when the acoustical properties of
the reproduction room fluctuate.
\par
Another category of ``active control'' soundfield reproduction methods
is based on the application of the Kirchhoff-Helmholtz integral
formula \cite{Ise,Takane,Betlehem,Gauthier}. In particular, the
Boundary Pressure Control technique (BPC) \cite{Takane} is based on
the assumption that the acoustic pressure inside a given volume
depends only on the pressure on the boundary surface, excepted at
certain frequency values. Secondary sources generate a soundfield
which is measured by microphones placed on the boundary surface of the
volume, and then compared with the soundfield to be reproduced. BPC
can be used to reproduce a given soundfield over a spatially extended
area free of microphones, and enables to accurately compensate for the
room reflections using adaptive filtering methods. This approach
therefore seems to be appropriate for reproducing low-frequency
soundfields.
\par
In this paper, the results of a feasibility study on the application
of BPC to low-frequency soundfield reproduction are presented. In
order to quantify the performances that can be expected when this
strategy is used, application of BPC to active noise control in free
field was tested, both numerically and experimentally. The reason why
this has not been done before is probably that the implementation of
this sound control strategy requires the use of many sensors and
actuators, which have to be managed by one or several high-performance
multichannel electronic controllers. Although the results presented in
this paper have been obtained in the context of active noise control,
similar performances could probably be obtained in that of soundfield
reproduction, since cancelling a noise and reproducing it constitute
basically the same physical task. Besides, the same hardware and
algorithms can be used in both tasks with very few changes
\cite{Nelson_jaes}.
\par
In the paper, the theory of BPC and its practical limitations are
presented in section \ref{theory}. Numerical BPC simulations in the
time and frequency domains are presented in section \ref{simulations}.
The experimental results obtained after real-time implementation of
BPC are given in section \ref{experiment}. An excellent agreement was
observed between the results of the simulations and of the
measurements, which make possible in the end the statement of
practical design rules for sound reproduction through BPC.

\section{Active control of sound using the Boundary Pressure Control
  method}
\label{theory}

\subsection{Kirchhoff--Helmholtz equation}
The Boundary Pressure Control technique involves the integral
representation of the acoustic field. As shown in Fig.~\ref{kirchfig},
let $\Omega$ denote a volume in space and $\Sigma$ its boundary
surface. If there is no acoustic source inside $\Omega$, then the
pressure at any point $\mathrm{\bf{r}}_{\Omega}$ in $\Omega$ (but not
on $\Sigma$) can be written as \cite{Bruneau}:
\begin{equation}
  \label{kircheq}
  p(\mathrm{\bf{r}}_{\Omega})=\int \!\!\! \int_{\Sigma} 
  \left( G(\mathrm{\bf{r}}_{\Omega},\mathrm{\bf{r}}_{\Sigma}) 
    \frac{\partial p(\mathrm{\bf{r}}_{\Sigma})}{\partial \mathrm{\bf{n}}_{\Sigma}} 
    - p(\mathrm{\bf{r}}_{\Sigma}) 
    \frac{\partial G(\mathrm{\bf{r}}_{\Omega},\mathrm{\bf{r}}_{\Sigma})}
    {\partial \mathrm{\bf{n}}_{\Sigma}} \right) \mathrm{d\bf{r}}_{\Sigma}
\end{equation}
where $G$ is the Green's function in free space, and
$\mathrm{\bf{n}}_{\Sigma}$ the unit vector normal to the surface.
This integral representation of the acoustic pressure, known as the
Kirchhoff--Helmholtz equation, shows that the acoustic pressure
measured inside a volume depends only on the pressure and its normal
derivative measured over the whole surface enclosing the volume.
Moreover, for a point $\mathrm{\bf{r}}_{\Omega}$ that lies on the
boundary $\Sigma$, it can be shown that \cite{Bruneau}:
\begin{equation}
  \label{kircheq2}
  \frac{1}{2} p(\mathrm{\bf{r}}_{\Sigma}^{0})=\int \!\!\! \int_{\Sigma} 
  \left( G(\mathrm{\bf{r}}_{\Sigma}^{0},\mathrm{\bf{r}}_{\Sigma}) 
    \frac{\partial p(\mathrm{\bf{r}}_{\Sigma})}{\partial \mathrm{\bf{n}}_{\Sigma}} 
    - p(\mathrm{\bf{r}}_{\Sigma}) 
    \frac{\partial G(\mathrm{\bf{r}}_{\Sigma}^{0},\mathrm{\bf{r}}_{\Sigma})}
    {\partial \mathrm{\bf{n}}_{\Sigma}} \right) \mathrm{d\bf{r}}_{\Sigma}
\end{equation}
In addition to Eq.~(\ref{kircheq}), Eq.~(\ref{kircheq2}) means that
there exists a linear relation between the pressure and its normal
derivative, both measured on $\Sigma$. The pressure inside $\Omega$
therefore depends only on the pressure measured on its boundary
surface. The general idea underlying the Boundary Pressure Control
technique is that one needs to impose the appropriate pressure value
only over the whole boundary surface to obtain the required soundfield
anywhere inside the volume. To reproduce of a given soundfield, the
method consists in: 1. recording the acoustic pressure over the whole
surface of the volume; 2. reproducing the same pressure values at the
points where they were recorded. Another way of interpreting
Eqs.~(\ref{kircheq}) and (\ref{kircheq2}) means that BPC can be used
for active noise control: with these equations, it is only necessary
to cancel the acoustic pressure only over the whole boundary surface
to cancel it anywhere inside the volume.
   
\subsection{Practical limitations of BPC}
Unfortunately, the use of the BPC technique has two serious
limitations.  The first one, which is commonly known as spatial
aliasing, is due to the spatial undersampling of the surface
controlled.  Eqs.~(\ref{kircheq}) and (\ref{kircheq2}) are integral
representation of the soundfield, which involve summing up the values
of the pressure and its normal derivative on a continuous distribution
of points over the whole surface under consideration. In practice,
this would involve controlling the soundfield at an infinite number of
points, which is of course impossible. Actually, in the practical
implementation of either a soundfield reproduction or an active noise
control system, it will not be possible to monitor the sound at more
that a few dozen points, and the performances of the system will
therefore decrease when the frequency increases
\cite{Kirkeby_1,ActiveControlOfSound}.
\par
The second physical limitation results from the mathematical
properties of Eq.~(\ref{kircheq2}). Assuming that the pressure is
equal to zero all over $\Sigma$, Eq.~(\ref{kircheq2}) becomes
\begin{equation}
  \label{dirichlet}
  \int \!\!\! \int_{\Sigma} 
  \left( G(\mathrm{\bf{r}}_{\Sigma}^{0},\mathrm{\bf{r}}_{\Sigma}) 
    \frac{\partial P(\mathrm{\bf{r}}_{\Sigma})}{\partial \mathrm{\bf{n}}_{\Sigma}} 
    - P(\mathrm{\bf{r}}_{\Sigma}) 
    \frac{\partial G(\mathrm{\bf{r}}_{\Sigma}^{0},\mathrm{\bf{r}}_{\Sigma})}
    {\partial \mathrm{\bf{n}}_{\Sigma}} \right) \mathrm{d\bf{r}}_{\Sigma}=0
\end{equation}
Finding the solutions to Eq.~(\ref{dirichlet}) is known as the
interior Dirichlet problem. Eq.~(\ref{dirichlet}) has an infinite
number of solutions when the frequency is equal to some eigenvalues
depending on the shape of the surface \cite{Kleinmann}. For example,
in the case of a rectangular parallelepiped, the corresponding
frequency values are given by
\begin{equation}
  \label{para_eigen_freq_eq}
  f_{\mathrm{lmn}} = \frac{c}{2}  {\sqrt{
      {\left( \frac{l}{L_{\mathrm{x}}} \right) }^{2} +  
      {\left( \frac{m}{L_{\mathrm{y}}} \right) }^{2} +  
      {\left( \frac{n}{L_{\mathrm{z}}} \right) }^{2} }}
\end{equation}
where $l$, $m$ and $n$ are strictly positive integers and
$L_{\mathrm{x}}$, $L_{\mathrm{y}}$ and $L_{\mathrm{z}}$ are the
dimensions of the parallelepiped. In the case of a spherical volume,
the characteristic wave numbers correspond to the zeros of the
spherical Bessel functions. In particular, the values
\begin{equation}
  \label{sphere_eigen_freq_eq}
  f_{\mathrm{k}} = k \frac{c}{2r}
\end{equation}
are eigenfrequencies of the problem, where $k$ is a positive integer,
$r$ the radius of the sphere and $c$ the celerity of sound. It is easy
to show that Eq.~(\ref{kircheq2}) also has an infinite number of
solutions at these frequencies. In practice, this means that if the
acoustic pressure is made to be equal to the appropriate value all
over $\Sigma$, it will not necessarily be equal to the appropriate
value inside $\Omega$ if the frequency is equal to one of the
eigenfrequencies of the interior Dirichlet problem related to
$\Omega$.
\par
In parts \ref{simulations} and \ref{experiment}, the feasibility of
soundfield reproduction using BPC is studied via simulations and
active noise cancellation experiments inside a sphere. The sphere was
chosen because, among the various shapes with the same capacity $V$,
the sphere is the shape with the lowest surface value. When performing
active noise control inside a volume with capacity $V$ using a BPC
strategy, the sphere therefore gives the best spatial discretization
of the boundary surface with a given number of minimisation
microphones. In addition, the sphere is the most regular shape
possible: the resonances resulting from the singularity in the
Dirichlet problem are therefore expected to be maximum in this case,
so that they can be easily detected. As far as the authors know, these
resonances have never been studied experimentally in the field of
active noise control or sound reproduction, because this requires the
use of a sufficiently large number of transducers, depending on the
shape of the volume in which the control is to be carried out.
Preliminary simulations showed that in the case of a sphere, about
$16$~microphones are required, whereas in the case of a cube, at least
$50$~microphones are required, which is a heavy constraint for active
noise control implementations.

\subsection{From active noise control to soundfield reproduction}
\label{anc2sr}
Let us consider an active noise control setup, including a number of
primary sources, secondary sources and minimisation microphones. In
the frequency domain, each acoustic path between a source and a
microphone is entirely described by a complex scalar. If
$\mathrm{\bf{G}}$ denotes the matrix of the acoustic paths between the
secondary sources and the minimisation microphones, and
$\mathrm{\bf{p}}_{0}$ the vector of the primary pressures measured at
the microphones, then the vector of optimal command signals for noise
cancellation with the secondary sources, regardless of any causality
or feasibility constraint, is given by:
\begin{equation}
  \label{command_noreg}
  \mathrm{\bf{u}}=-{\mathrm{\bf{G}}}^{-1}{\mathrm{\bf{p}}}_{0}
\end{equation}
If Tikhonov regularization \cite{Tikhonov} is introduced into the
matrix inversion process, which corresponds to adding an effort
weighting to the command vector computation
\cite{SignalProcessingForActiveControl}, Eq.~(\ref{command_noreg})
becomes
\begin{equation}
  \label{command_reg}
  \mathrm{\bf{u}}=-{(\mathrm{\bf{G}}^{\mathrm{H}}\mathrm{\bf{G}} 
    + \beta I )}^{-1}\mathrm{\bf{G}}^{\mathrm{H}}{\mathrm{\bf{p}}}_{0}
\end{equation}
where $\beta$ is the regularization coefficient. Such a coefficient
usually helps in widening the minimisation area around the microphones
because with regularization the noise minimisation problem at a finite
number of microphones fits better the underlying global minimisation
problem at an infinite number of locations. Therefore, if
$\mathrm{\bf{H}}$ denotes the matrix of acoustic paths between the
secondary sources and some observation points, the total pressure
$\mathrm{\bf{p}}_{\mathrm{tot}}$ measured at these points when the
control is on is given by the formula
\begin{eqnarray}
  \label{total_pres}
  {\mathrm{\bf{p}}}_{\mathrm{tot}} & = & {\mathrm{\bf{p}}}_{\mathrm{pri}} 
  + {\mathrm{\bf{p}}}_{\mathrm{sec}}\nonumber \\
  & = & {\mathrm{\bf{p}}}_{\mathrm{pri}} 
  + {\mathrm{\bf{H}}}\mathrm{\bf{u}}\nonumber \\
  & = & {\mathrm{\bf{p}}}_{\mathrm{pri}}
  -\mathrm{\bf{H}}{(\mathrm{\bf{G}}^{\mathrm{H}}\mathrm{\bf{G}} 
    + \beta I )}^{-1}\mathrm{\bf{G}}^{\mathrm{H}}{\mathrm{\bf{p}}}_{0}
\end{eqnarray}
where ${\mathrm{\bf{p}}}_{\mathrm{pri}}$ is the vector of pressures
when the control is off, and ${\mathrm{\bf{p}}}_{\mathrm{sec}}$
denotes the vector of secondary pressures, both measured at the
observation points. The mean attenuation obtained on $N$ observation
points ($\mathrm{\bf{x}}_{1}$, $\mathrm{\bf{x}}_{2}$, ...
$\mathrm{\bf{x}}_{\mathrm{N}}$) can then be written:
\begin{equation}
  \label{att_ave}
  A=\frac{1}{N} \sum\limits_{k=1}^{N} 20{log}_{10}
  \left| \frac{{\mathrm{\bf{p}}}_{pri}(\mathrm{\bf{x}}_{\mathrm{k}})}
    {{\mathrm{\bf{p}}}_{tot}(\mathrm{\bf{x}}_{\mathrm{k}})} \right|
\end{equation}
\par
Let us now consider the case where the same arrangement of sources and
microphones is used for soundfield reproduction purposes. The task now
consists in generating a soundfield with the secondary sources which
is as similar as possible to the primary soundfield at the
minimisation microphones. The vector of optimal command signals with
regularization is given here by:
\begin{eqnarray}
  \mathrm{\bf{u}'} & = & {(\mathrm{\bf{G}}^{\mathrm{H}}\mathrm{\bf{G}} 
    + \beta I )}^{-1}\mathrm{\bf{G}}^{\mathrm{H}}{\mathrm{\bf{p}}}_{0}
  \nonumber \\
  & = & -\mathrm{\bf{u}}
\end{eqnarray}
Hence, the vector of the reproduced pressures, which is also the
vector of the secondary pressures measured at the observation points,
is given by:
\begin{equation}
  \label{machin}
  {\mathrm{\bf{p}}}_{\mathrm{sec}}'=-{\mathrm{\bf{p}}}_{\mathrm{sec}}
\end{equation}
Whereas the attenuation is the most appropriate criterion for
assessing the noise control performances, the quality of the
reproduction can be measured in terms of relative error:
\begin{equation}
  \label{truc}
  E=\frac{1}{N} \sum\limits_{k=1}^{N} 20{log}_{10}
  \left| \frac{{\mathrm{\bf{p}}}_{\mathrm{pri}}(\mathrm{\bf{x}}_{\mathrm{k}})
      - {\mathrm{\bf{p}}}_{\mathrm{sec}}'(\mathrm{\bf{x}}_{\mathrm{k}})}
    {{\mathrm{\bf{p}}}_{\mathrm{pri}}(\mathrm{\bf{x}}_{\mathrm{k}})} \right|
\end{equation}
Replacing ${\mathrm{\bf{p}}}_{\mathrm{sec}}'$ in Eq. \ref{truc} by the
value given in Eq.~\ref{machin}, we obtain:
\begin{eqnarray}
  E & = & \frac{1}{N} \sum\limits_{k=1}^{N} 20{log}_{10}
  \left| \frac{{\mathrm{\bf{p}}}_{\mathrm{pri}}(\mathrm{\bf{x}}_{\mathrm{k}})
      + {\mathrm{\bf{p}}}_{\mathrm{sec}}(\mathrm{\bf{x}}_{\mathrm{k}})}
    {{\mathrm{\bf{p}}}_{\mathrm{pri}}(\mathrm{\bf{x}}_{\mathrm{k}})} \right| 
  \nonumber \\
  & = &  \frac{1}{N} \sum\limits_{k=1}^{N} 20{log}_{10}
  \left| \frac{{\mathrm{\bf{p}}}_{\mathrm{tot}}(\mathrm{\bf{x}}_{\mathrm{k}})}
    {{\mathrm{\bf{p}}}_{\mathrm{pri}}(\mathrm{\bf{x}}_{\mathrm{k}})} \right| 
  \nonumber \\
  & = & -A
\end{eqnarray}
The reconstruction error is therefore the opposite of the attenuation
obtained in the case of active noise cancellation. In other words, if
the active noise control setup attenuates a given noise of $40$~dB, it
will be able to reproduce the primary soundfield pressure with a
relative error of $-40$~dB, e.g. $1$~\%.
\par
Assuming that the acoustic paths are the same in both cases, the
optimum active noise control and soundfield reproduction performances
will be similar in the frequency domain, which is quite natural, since
in both cases, the performances depend only on the inversion of the
matrix of secondary acoustic paths. For the same reasons, the
performances of noise cancellation and reproduction will also be
similar using adaptive algorithms in the time domain. It is therefore
possible to assess the performances of a soundfield reproduction setup
using it as an active noise control setup, and vice-versa.

\section{Numerical simulations}
\label{simulations}

\subsection{The setup used in simulations}
In this section, the results of preliminary numerical simulations
performed in the case of the $30\times30$ multi-channel active noise
control of a spherical volume are presented. It was proposed here to
simulate the behaviour of an active noise control setup in order to
compare the results with those obtained experimentally
(part~\ref{experiment}). However, it is worth noting that the results
would be exactly the same if the setup simulated was a soundfield
reproduction system, as shown in section~\ref{anc2sr}. Firstly, the
system was simulated in the frequency domain in order to determine the
optimum predictable attenuation. The results obtained were most
encouraging, although the two limitations mentioned in part
\ref{theory} dramatically restricted the noise cancellation
performances inside the volume. Secondly, the system was simulated in
the time domain in order to assess the performances of a real system
involving a Filtered-X Least Mean Square (FXLMS) time-domain
algorithm. Two types of primary fields were tested: pure tones, and
broadband noises. The results were predictably less satisfactory here
than in the frequency-domain case, because it was difficult to make
the algorithm converge. An interesting finding which emerged was the
fact that the problem of Dirichlet resonances is a serious weakness of
the system even in the case of a broadband primary soundfield.
\par
Figs.~\ref{simul__sphere_mic_fig} and \ref{simul__setup_fig} show the
setup used in the numerical simulations. It was composed of one
virtual primary source, $30$ virtual secondary sources, and $30$
minimisation microphones. As shown in
Fig.~\ref{simul__sphere_mic_fig}, the virtual microphones were
distributed on the surface of a sphere with a diameter of $30$~cm,
along $8$ arcs of a circle, as far from each other as possible. The
positions of the secondary sources were the homothetical images of the
microphone positions on a sphere with a radius of $60$~cm radius. The
primary source was located on the same horizontal plane as the center
of the sphere, $4$~m away from it. In the computations, all the
transducers were assumed to be perfect monopoles in free field:
therefore, the impulse responses of each source measured at each
microphone were perfect impulses with a time-delay of $\frac{r}{c}$
seconds, attenuated by a factor $r$, where $r$ denotes the distance
between the transducers.

\subsection{Frequency-domain simulations}
Frequency-domain simulations were carried out using the optimal
control formulation given in section \ref{anc2sr}.
Fig.~\ref{simul__freq__att_moy_volu} shows the average attenuation
obtained with a mesh consisting of $160$~points regularly spaced
inside the sphere as a function of the frequency, without any
regularisation procedure ($\beta=0$ in Eq.~\ref{command_reg}). This
figure illustrates the two physical limitations mentioned in part
\ref{theory}. In a first approximation, it can be seen that the
average attenuation decreased linearly from $0$ to $1000$~Hz, by
approximately $6$~dB per Hz. This regular decrease in the control
performances was due to the spatial undersampling: the higher the
frequency, the less satisfactory the discretization of the soundfield
over the boundary surface and the control performance became. In
addition, the control performances dropped dramatically when the
frequency of the primary soundfield approached the eigenfrequencies of
the interior Dirichlet problem. The attenuation even fell to less than
$-30$~dB, which means that the interior pressure level present when
the control was on was more than ten times that occurring when control
was off. This was because only the acoustic pressure was controlled,
although both the pressure and its normal derivative should be
controlled: the sound pressure can have any value inside the sphere,
even if the control is highly efficient at the level of the
minimisation microphones.
\par
Fig.~\ref{simul__freq__att_moy_volu_reg} shows the results obtained in
the case where $\beta$ was taken to be equal to $0.9$. It shows that
it is possible to prevent the pressure attenuation from being negative
at the eigenfrequencies of the Dirichlet problem by regularizing the
matrix before it is inverted. This improvement is achieved with no
great loss in the control performances inside the sphere, as shown in
Fig.~\ref{simul__freq__att_moy_volu_comp}. Another advantage of the
regularization procedure is that it decreases the amplitude of the
command vector. This suggests that in the time domain, it might be
useful to add a leakage term to the control algorithm
\cite{SignalProcessingForActiveControl}, which is similar to
performing Tikhonov regularization in the frequency domain and can
usefully increase the stability of an active noise control system of
this kind. Note that in the case of a regularized matrix inversion
process, the attenuation observed inside the sphere can be greater
than that observed at the minimisation microphones.

\subsection{Time-domain simulations}
In the time domain, the acoustic paths between each source and each
microphone are described by impulse responses. In the case of perfect
point sources under free-field conditions, the impulse responses are
time-delayed, attenuated pulses. However, since the time delays are
mostly not exact numbers of sampling periods, the impulse responses
need to be approximated for discrete-time simulation purposes. The
accuracy of the approximation depends strongly on the sampling
frequency. In this study the frequency chosen was $8192$~Hz, which has
been found to suffice in view of the dimensions of the setup: the
shortest distance between two transducers was $30$~cm, which
corresponds to a delay of approximately $7$~samples at this frequency.
The transfer functions were first computed in the frequency domain in
$8192$ frequency bins, and they were then converted into impulse
responses by performing an Inverse Fast Fourier Transform (IFFT) and
truncated to their first $256$ coefficients. The impulse responses
obtained were finally used in a program simulating a multichannel
FXLMS algorithm in the time domain.  The same formulation as in
\cite{SignalProcessingForActiveControl} was used to adapt the filter
coefficients:
\begin{equation}
  \label{FXLMS}
  \mathrm{\bf{w}}(n+1)=\mathrm{\bf{w}}(n)
  -\alpha \left( \hat{\mathrm{\bf{R}}}^{\mathrm{T}}(n) \mathrm{\bf{e}}(n) 
    +{\beta}' \mathrm{\bf{w}}(n) \right) 
\end{equation}
where $\mathrm{\bf{w}}(n)$ denotes the vector of the minimisation
filter coefficients at the $n$th sample time,
$\hat{\mathrm{\bf{R}}}(n)$ denotes the matrix of the estimated
filtered reference signals, $\mathrm{\bf{e}}(n)$ denotes the vector of
the error signals, $\alpha$ denotes the convergence coefficient, and
${\beta}'$ is a regularization coefficient. Note that the primary
source command signal was used as the reference signal by the FXLMS
algorithm. Two types of simulation were carried out, corresponding to
two types of primary signals: firstly, a pure tone signal at various
frequencies; and secondly, a broadband signal with a frequency range
of $0$--$900$~Hz.
\par
In the case of pure tone primary signals, the frequency was increased
step by step: at each frequency value, the algorithm was let to
converge during a few thousand samples, and the value of the
attenuation was then averaged based on the last thousand samples and
saved. Note that the $\alpha$ convergence coefficient was the same at
each frequency value. As in the experimental case, the length of the
estimated secondary paths was $200$~coefficients and the length of the
minimisation filters was $10$~coefficients.
Fig.~\ref{simul__temp__sweptsine_att_moy} shows the pressure
attenuation obtained with pure tone primary signals from $200$ to
$1000$~Hz at the minimisation microphones, compared to the averaged
pressure attenuation inside the sphere. The results obtained here were
very similar to those obtained in the frequency domain in the
regularized case throughout the frequency band, but the control was
less efficient in the time domain. The attenuation is not shown here
at frequencies inferior to $200$~Hz, because the algorithm had
difficulty in converging at lower frequencies, and the computation
time required for the algorithm to converge with a smaller value of
the $\alpha$ coefficient would have been too long. This convergence
problem results from the ill-conditioning of the matrix of secondary
paths at low frequencies \cite{ActiveControlOfSound}, which again
confirms the importance of using leakage in systems of this kind.
\par
In the case of a broadband primary signal, the length of the
minimisation filters and estimated secondary paths were both set at
$90$~coefficients. The algorithm was made to converge during several
dozen thousands of samples, and the minimisation filters obtained were
then used to compute in the frequency domain the pressure attenuation
occurring inside the sphere and at the minimisation microphones.
Fig.~\ref{simul__temp__broadband} shows the control performances
obtained using this method with leakage in the case of a $0$--$900$~Hz
white primary noise. Similar findings to those made in the case of
pure tone signals can be obtained here: although the control
performances observed at the minimisation microphones are good
throughout the frequency range of the primary signal, the average
pressure attenuation obtained inside the volume becomes poor at
frequencies approaching the eigenvalues of the Dirichlet problem. Note
that for frequencies below $500$~Hz, the performances obtained inside
the sphere were more satisfactory than those obtained at the
minimisation microphones.

\subsection{Conclusions}
Several important conclusions can be drawn from this preliminary
study. First, all the simulations presented here showed how
efficiently the system controls low-frequency noise. The control
performances were satisfactory throughout the volume at frequencies
below $500$~Hz in all the cases tested. At this frequency, the
wavelength is $68$~cm, which is approximately equal to the diameter of
the sphere: the control is therefore no longer {\em local} and
actually includes the whole volume. Secondly, the inaccuracy of the
spatial sampling and the non-uniqueness of the interior Dirichlet
problem seem to be two important limitations of this control method,
including for control of broadband noises. Thirdly, leakage appears to
be a useful means of reducing the resonance problems which occur when
the frequency tends to the eigenvalues of the Dirichlet problem.

\section{Experiments}
\label{experiment}

\subsection{Experimental setup}
Although the aim of this study was to investigate the feasibility of
soundfield reproduction using BPC, we decided to assess the setup
performances by performing active noise control experiments using the
same strategy. The controllers used here had already been programmed
with an active noise control algorithm, and the behaviour of the setup
was expected to be similar in the case of both active noise
cancellation soundfield reproduction, as shown in part \ref{anc2sr}.
\par
The setup used in these experiments is shown in
Fig.~\ref{expe__photo_1}. Thirty minimisation microphones were
distributed over the whole surface of a sphere with a diameter of
$60$~cm, as in the case of the setup modelled in the numerical
simulations. The primary soundfield was emitted by a primary
loudspeaker placed at a few meters from the microphone sphere, and was
controlled by thirty secondary sources distributed over the surface of
a sphere with a diameter of $170$~cm, so that the distance between
each secondary source and the nearest minimisation microphone was
approximately $30$~cm. Secondary sources and minimisation microphones
were numbered so that microphone $1$ was the nearest microphone to
source $1$, and so on. In addition to the minimisation microphones,
two other microphones were placed inside the microphone sphere in
order to measure the control performances inside the sphere: one was
in the center, and the other one was approximately mid-way between the
center of the sphere and its surface. As in the numerical simulations,
the control performances were measured in the case of two types of
primary noises: pure tone noise, and broadband noise.
\par
In the case of the pure tone noise, the secondary source command
signals were computed by a $32\times32$ multichannel controller
designed for active control of tonal disturbances. The secondary paths
identified were saved into FIR filters with $200$ coefficients and the
length of the minimisation FIR filters was arbitrarily chosen as $10$
coefficients, although $2$~coefficient may have sufficed in theory.
The frequency of the primary sound was increased by $5$~Hz every
$16$~s from $200$ to $700$~Hz: the algorithm was let to converge
during the first $8$~s and the acoustic pressure was measured during
the last $8$~s. The convergence coefficient $\alpha$ remained constant
throughout the measurement process. In both the measurement process
and the control process, the sampling frequency was $2048$~Hz.
\par
In the case of broadband noise, however, secondary source command
signals were computed by two $16\times16$ multichannel LMA COMPARS
controllers \cite{Compars} programmed with an FXLMS algorithm, as
shown in Figs.~\ref{expe__photo_2} and \ref{expe__broadband__setup}.
The inputs to the first COMPARS were the pressure signals from
microphones $1$--$16$, and the outputs were the secondary sources
$1$--$16$ command signals; the second COMPARS received the signals
from microphones $17$--$30$, and computed the commands to be
transmitted to sources $17$--$30$. Note that the microphones and
sources were spatially distributed in two blocks in order to ensure
the simultaneous convergence of the both algorithms. The whole active
noise control setup, including the two controllers, can be viewed in
fact as a single FXLMS system with a block-diagonal matrix of
identified secondary paths: a sufficient condition for this system to
converge is that the matrix of real secondary paths is block-diagonal
dominant \cite{Leboucher}. Note also that the experiment was conducted
in a quasi-anechoic environment to facilitate the convergence of the
algorithms. If the system had been implemented in a more reverberant
room, the reflections from the walls would have increased the effects
of each secondary source on the distant minimisation microphones, and
the transfer matrix would have been less diagonally dominant. Lastly,
the electric signal fed to the primary source was used by both
controllers as a reference signal for the FXLMS algorithm. The
identified secondary paths and the control filters both consisted of
FIR filters with $90$ coefficients. As in the pure tone case, a
sampling frequency of $2048$~Hz was used. The primary noise was
therefore a white noise with a $0$ to $1024$~Hz spectrum. The
convergence coefficient ($\alpha$ in Eq. \ref{FXLMS}) was taken to be
approximately half of the value at which the algorithm began to
diverge and was the same in the two controllers.

\subsection{Results in the pure tone case}
Fig.~\ref{expe__sweptsine__att} shows the attenuation measured at the
minimisation and observation microphones in the case of pure tone
primary signals with frequency ranging from $100$ to $700$~Hz. In the
$200$--$500$~Hz frequency band, the attenuation values measured both
on the surface of the sphere and inside it were greater than $30$~dB,
which corresponds to a highly efficient control in the whole volume.
Beyond $500$~Hz, however, the attenuation dropped off dramatically at
the observation microphones, reaching a minimum value of only a few dB
at $580$~Hz, although it was still above $30$~dB at the minimisation
microphones. Note that the minimum attenuation was obtained when the
frequency ranged between $580$ and $590$~Hz, which is slightly above
the value of $566$~Hz expected for a sphere with a radius of $30$~cm.
However, this amount of frequency shift was more or less expected,
because the measured eigenfrequency corresponds to a radius of $29$~cm
and $1$~cm was approximately the accuracy of the minimisation
microphone localization. Note also that the attenuation decreased at
frequencies below $200$~Hz. This was due to two factors: first, the
secondary sources used for the noise control were small, and therefore
not very powerful at low frequencies; second, the convergence of the
algorithm was slow in this frequency range, probably because of the
ill-conditioning of the matrix of secondary paths. One interesting
result emerged when the attenuation curves of the two interior
microphones were compared: the attenuation measured at microphone~$1$
when the frequency was close to the first eigenfrequency of the sphere
is lower than that measured at microphone~$2$. This finding can be
explained by the fact that the first eigenmode of the sphere is
radial, the maximum pressure value being reached in the center of the
sphere, where microphone~$1$ was located.
\par
Good agreement was found to exist here between the experimental data
and the results of the simulations, as shown in
Fig.~\ref{expe__sweptsine__comp}. The value of the sphere radius was
set at $29$~cm in the simulation, in order to take into account the
shift of the first eigenfrequency observed experimentally: this
corresponds to the first eigenfrequency occurring at approximately
$585$~Hz. It can be seen from Fig.~\ref{expe__sweptsine__comp} that
even if the assumptions made in the numerical simulations (point
sources in free field) are not met in the experiment, the computations
give a very good idea of what occurs in practice. On the one hand, the
frequency-domain simulations of optimal control give an approximation
of the maximum attenuation values reached in the experiments. On the
other hand, the values obtained in the time-domain simulations were
very similar to those measured in practice, including the differences
in the attenuation between the two interior microphones. Note that the
minimum attenuation value was obtained at interior microphone~$1$,
which was placed in the center of the sphere, under both real and
simulated conditions.

\subsection{Results in the broadband case}
Fig.~\ref{expe__broadband__frf} shows the amplitude of the frequency
response measured between the primary source command signal and the
minimisation microphones in the case of a $0$--$1024$~Hz broadband
primary sound, with and without control. As was to be expected the
control performances were less satisfactory here than with pure tone
signals. However, the control was found to be highly efficient, since
the attenuation achieved was above $15$~dB in almost the whole
frequency range under consideration. Note that the poor control
performances measured at frequencies around $150$~Hz are due to the
insufficient electrical insulation between the components of the
experimental setup and the power supply.
\par
The control performances obtained inside the sphere are given in
Fig.~\ref{expe__broadband__att}. As in the case of pure tone signals,
the pressure attenuation achieved inside the sphere was greater than
the attenuation at the surface at frequencies below $500$~Hz. Below
this frequency, the setup is therefore able to efficiently cancel a
random noise anywhere in the sphere. Above this frequency, however,
the attenuation decreases inside the sphere, reaching a minimum value
of around $0$~dB at a frequency of $585$~Hz. Note that with interior
microphone~$1$, which is in the center of the sphere, the attenuation
is even negative between $570$ and $600$~Hz. The pressure measured at
the interior microphone~$1$ when the control is on is therefore
greater than the pressure measured when the control is off at these
frequency values. Note also that the magnitude of the resonance
occurring at around $585$~Hz at microphone~$1$ is greater than at
microphone~$2$. This confirms that the resonance induced at this
frequency corresponds to a radial eigenmode, which reaches a maximum
value in the center of the sphere.
\par
In Fig.~\ref{expe__broadband__comp}, the experimental results are
compared with those of the simulations. It can be seen that the
simulation does not model the experimental setup behaviour above
$600$~Hz. However, the agreement between the measured attenuation and
the numerically computed values is very good below this frequency. The
simulated attenuation values calculated at both the minimisation
microphones and the interior microphones are very similar to those
obtained in practice. Besides, as in the pure tone case, the
differences in noise attenuation at the interior microphones are well
reproduced, including the fact that around $585$~Hz the attenuation
reaches a minimum in the center of the sphere.

\subsection{Conclusions: interpretation in terms of soundfield
  reproduction}
The experimental results obtained with the present active noise
control setup can be interpreted in terms of soundfield reproduction,
as described in part \ref{anc2sr}. These results show that an
attenuation amounting to more than $30$~dB was obtained in the pure
tone signals at frequencies below $500$~Hz. The setup is therefore
able to reproduce pure tone soundfields in the same frequency range
with an error of less than $3$~\%, which is very accurate. In the case
of broadband noise, the results show that the reconstruction error can
reach about $10$~\% ($-20$~dB) in the $100-500$~Hz frequency range.
These results are most encouraging and show that BPC has considerable
potential for use as a low-frequency soundfield reproduction strategy.
\par
However, the negative attenuation observed when Dirichlet resonance
occurred suggests the presence of a reconstruction error grater than
$100$~\%, which means that a reproduction system based on the use of
BPC would be completely unable to reproduce soundfields at the
eigenfrequencies of the internal Dirichlet problem.

\section{General conclusions and perspectives}
\label{conclusion}
In this study, it was established that an active noise control setup
based on the Boundary Pressure Control technique can efficiently
cancel noise everywhere inside a volume. The existence of considerable
similarities between active noise control and soundfield reproduction
suggests that similar performances can be obtained when BPC is used to
reproduce soundfields. However, the results of both simulations and
experimental measurements show that spatial aliasing and the
non-uniqueness of the solution to the interior Dirichlet problem limit
the use of this strategy. In particular, resonance processes occurring
inside the volume when the frequency approaches the eigenfrequencies
of the Dirichlet problem completely prevent the system from cancelling
the primary noise at these frequencies. It was observed that in
practice the attenuation can even become negative when these
resonances occur. In the case of soundfield reproduction applications,
this means that the reconstruction error could be greater than 100~\%
at these frequencies. Besides, the data obtained here show that the
problem occurs in the case of both pure tone signals and broadband
noise signals. Thus, finding a means of reducing the drop in the
performances resulting from this problem without increasing the number
of transducers required would be an interesting goal for future
studies. Some solutions have already been proposed in \cite{Takane}
and \cite{Ise}, such as adding a minimisation microphone inside the
volume under consideration, but these solutions have not yet been
tested in practice, nor in 3D numerical simulations.
\par
This study was carried out in an almost anechoic environment, in order
to simplify the impulse responses which had to be compensated by the
controller. However, in order to reach high pressure levels at very
low frequencies required by a realistic reproduction of sonic booms,
the reproduction system has to be implemented in a room, and the
reflections occurring under these conditions cannot be neglected.
Longer minimisation filters will probably be required to achieve
comparable performances in a room, as the impulse responses of the
secondary sources will be longer under these conditions than in a
quasi-anechoic environment. The numerical simulation of an in-room
soundfield reproduction system has been previously carried out
\cite{active04} and the results suggest systems of this kind can
accurately reconstruct low-frequency wavefronts. The authors therefore
intend in the future to implement low-frequency soundfield
reproduction strategies using BPC in a specially designed reproduction
room.

% The Appendices part is started with the command \appendix; appendix
% sections are then done as normal sections
% \appendix

% \section{}
% \label{}

\newpage {\bf{List of figure captions}}
\par
\begin{figure}
  \vspace*{4cm} \centering \includegraphics{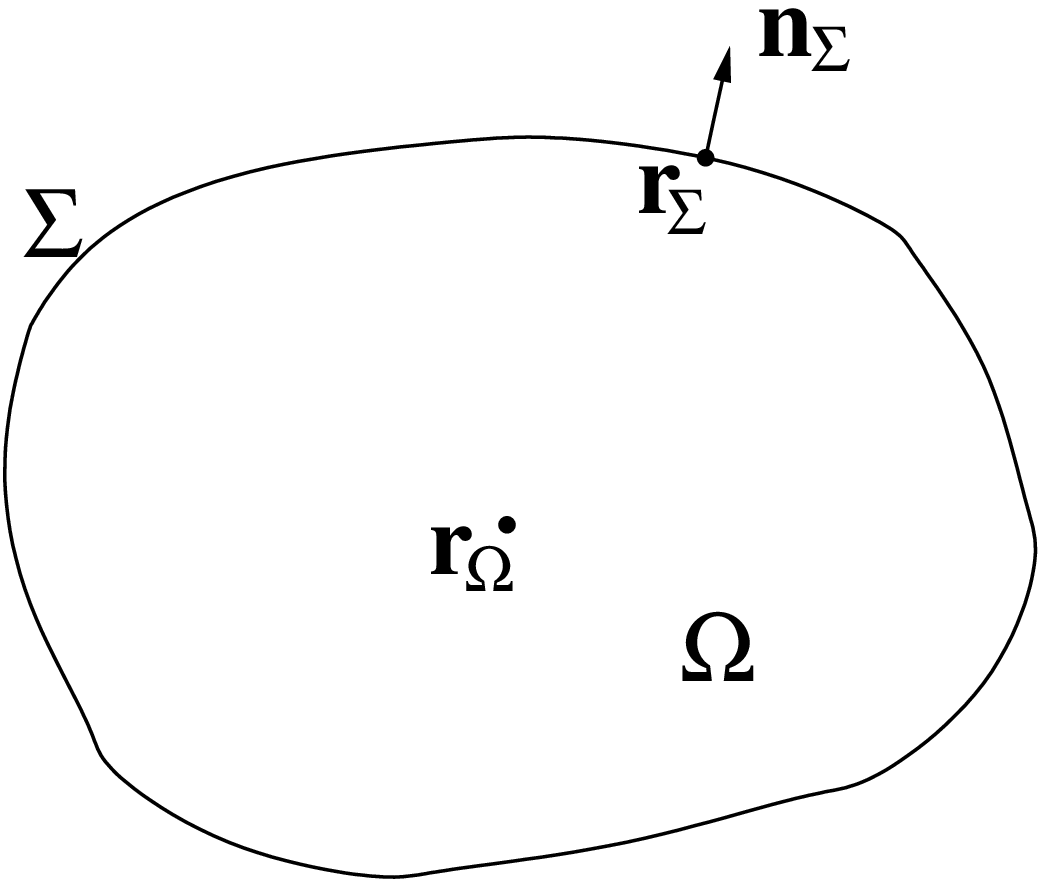}
  \vspace*{5cm}
  \caption{}
  \label{kirchfig}
\end{figure}
\begin{figure}
  \vspace*{4cm} \centering
  \includegraphics[width=12cm]{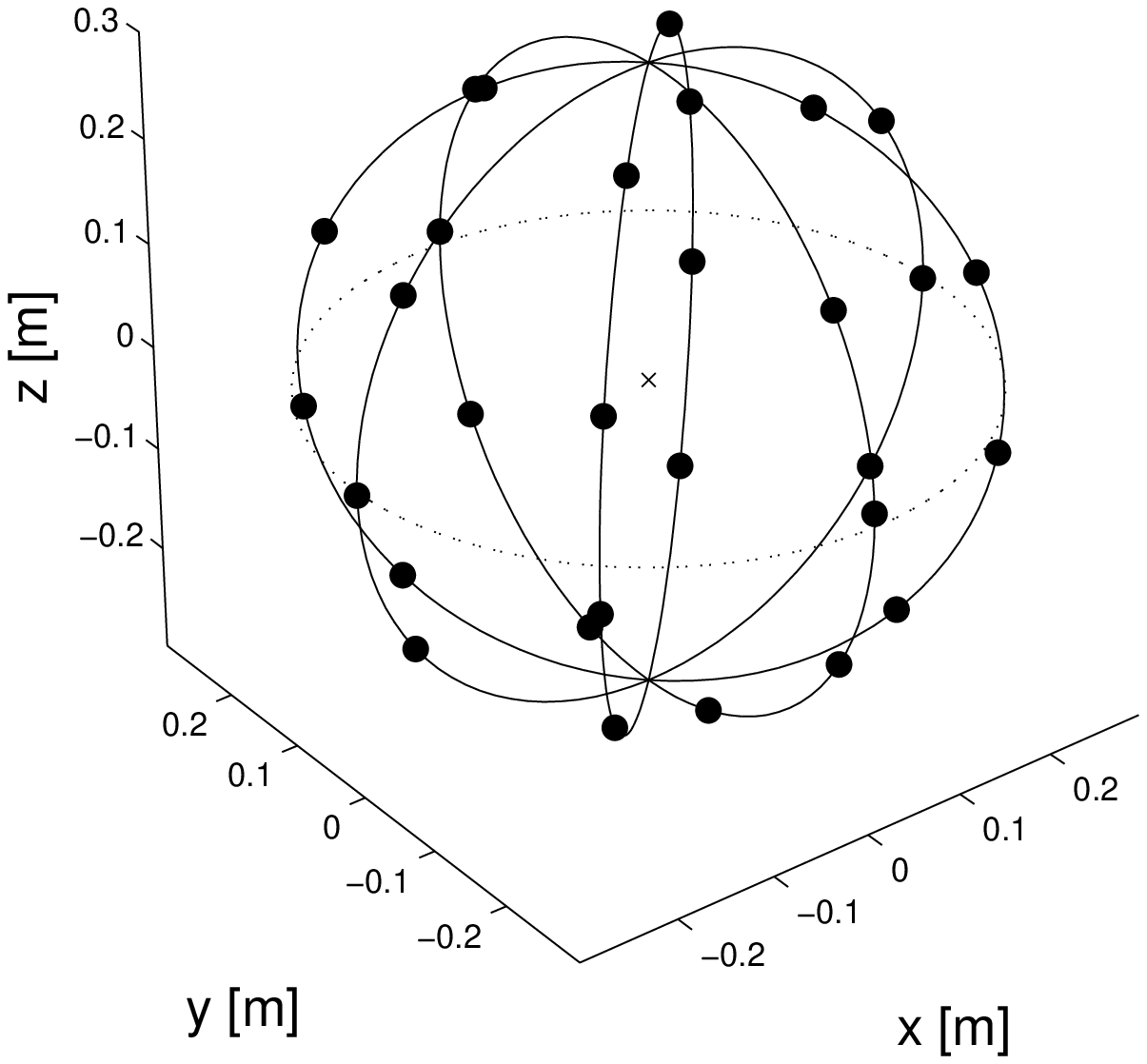}
  \vspace*{5cm}
  \caption{}
  \label{simul__sphere_mic_fig}
\end{figure}
\begin{figure}
  \vspace*{4cm} \centering
  \includegraphics{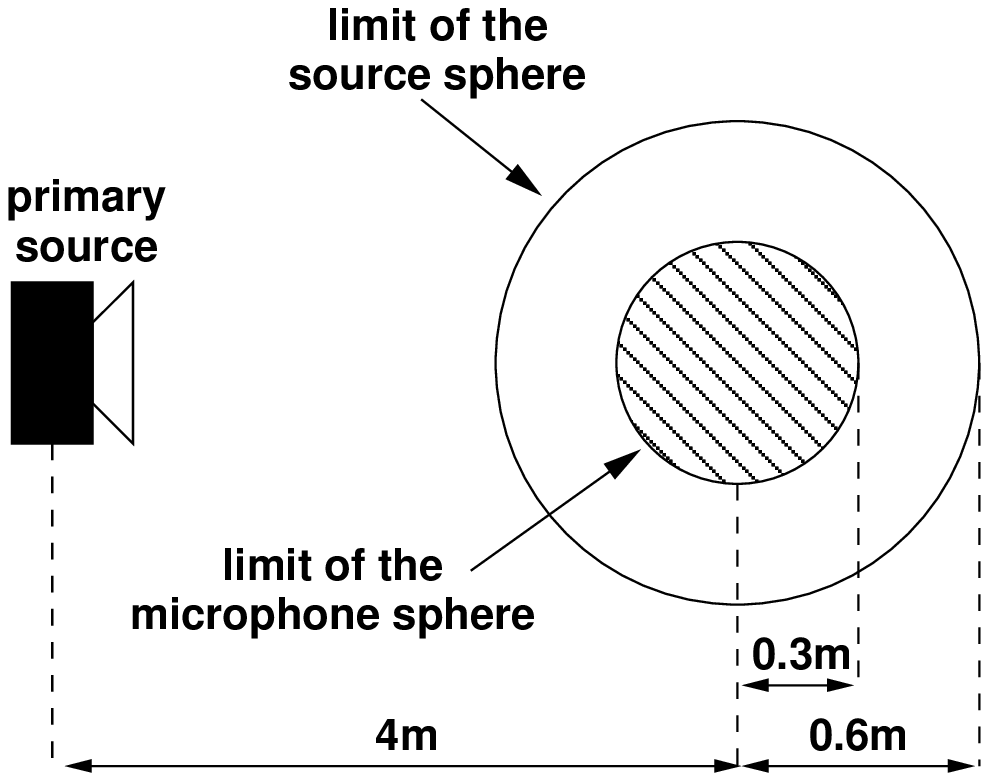} \vspace*{5cm}
  \caption{}
  \label{simul__setup_fig}
\end{figure}
\begin{figure}
  \vspace*{4cm} \centering
  \includegraphics[width=12cm]{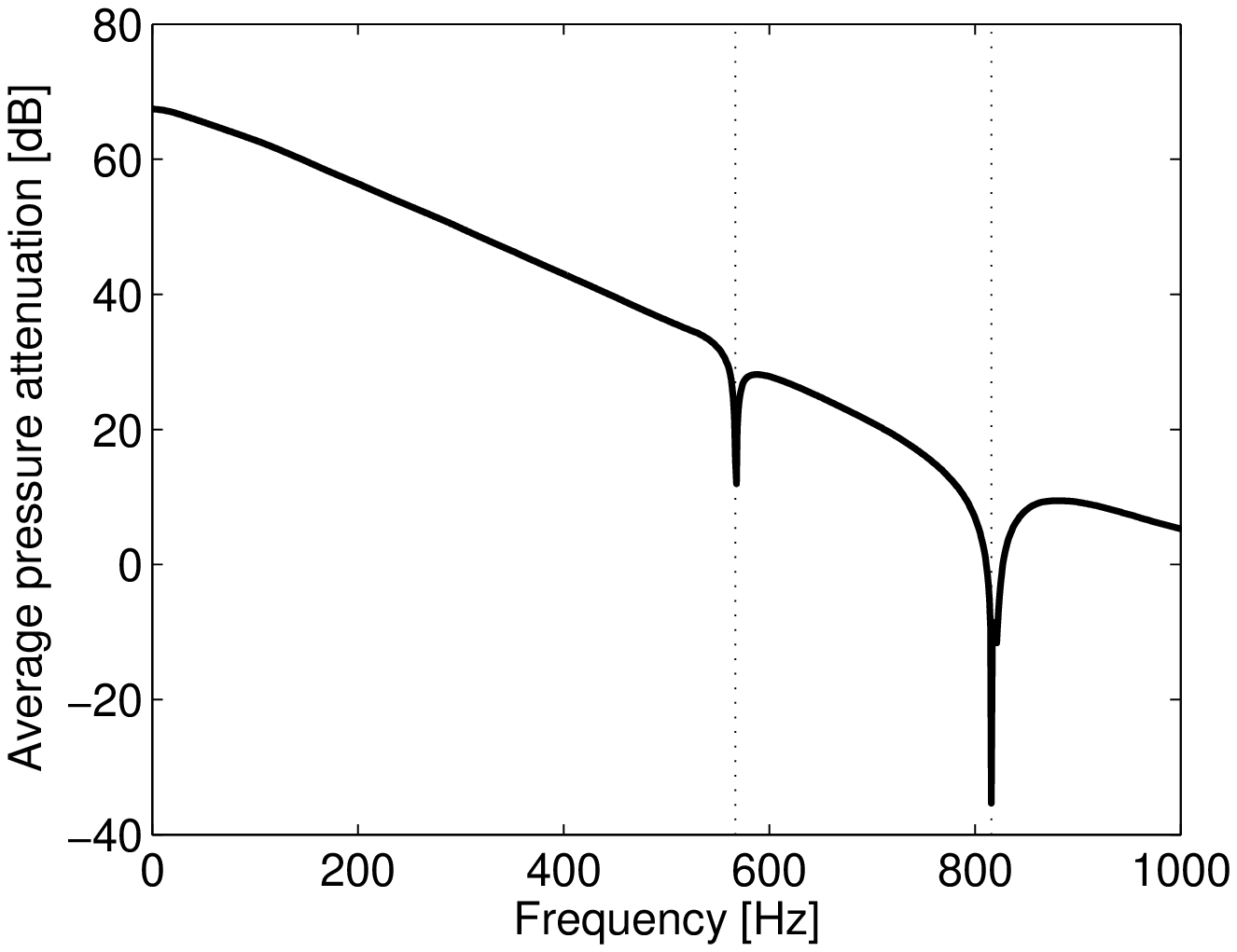}
  \vspace*{5cm}
  \caption{}
  \label{simul__freq__att_moy_volu}
\end{figure}
\begin{figure}
  \vspace*{4cm} \centering
  \includegraphics[width=12cm]{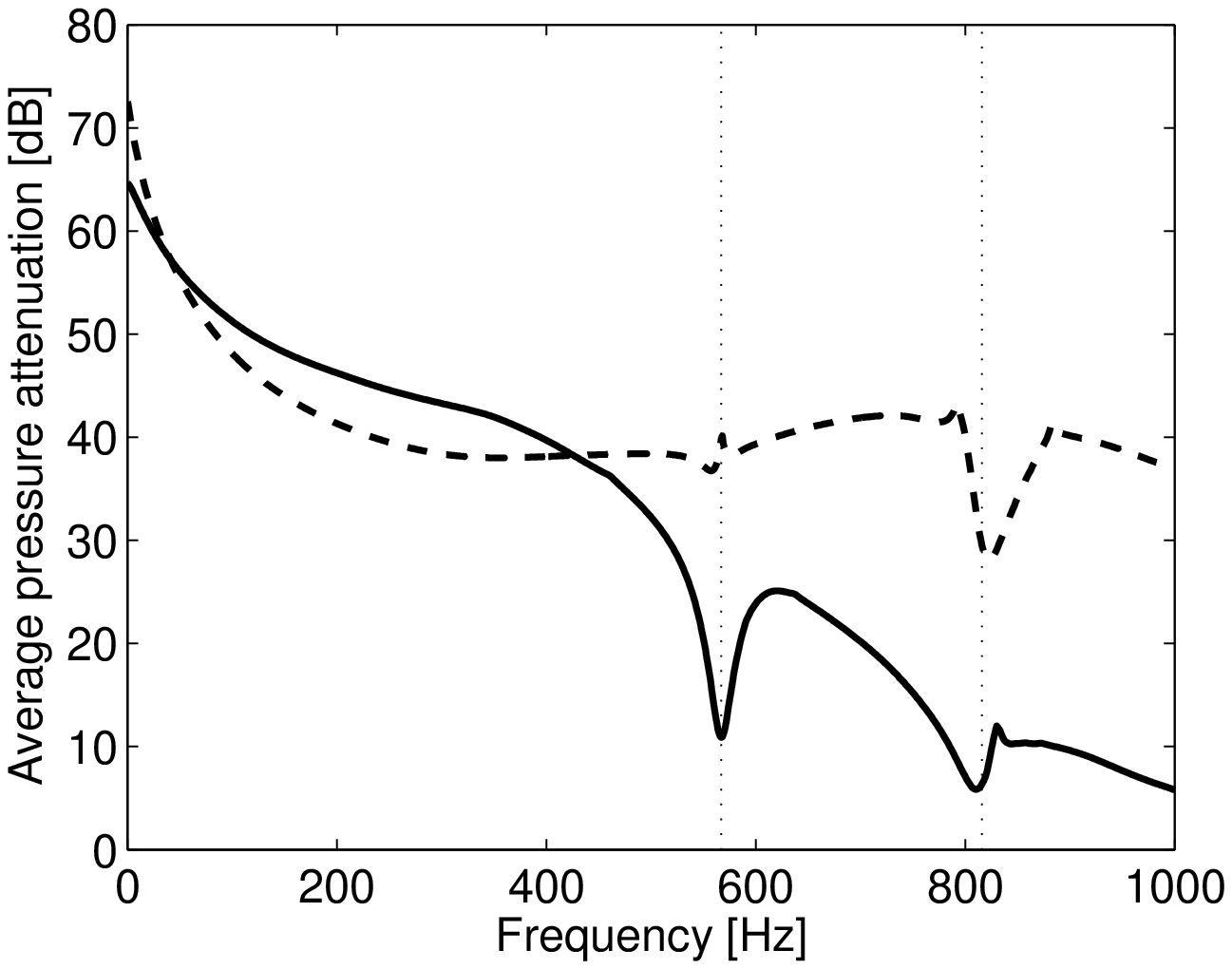}
  \vspace*{5cm}
  \caption{}
  \label{simul__freq__att_moy_volu_reg}
\end{figure}
\begin{figure}
  \vspace*{4cm} \centering
  \includegraphics[width=12cm]{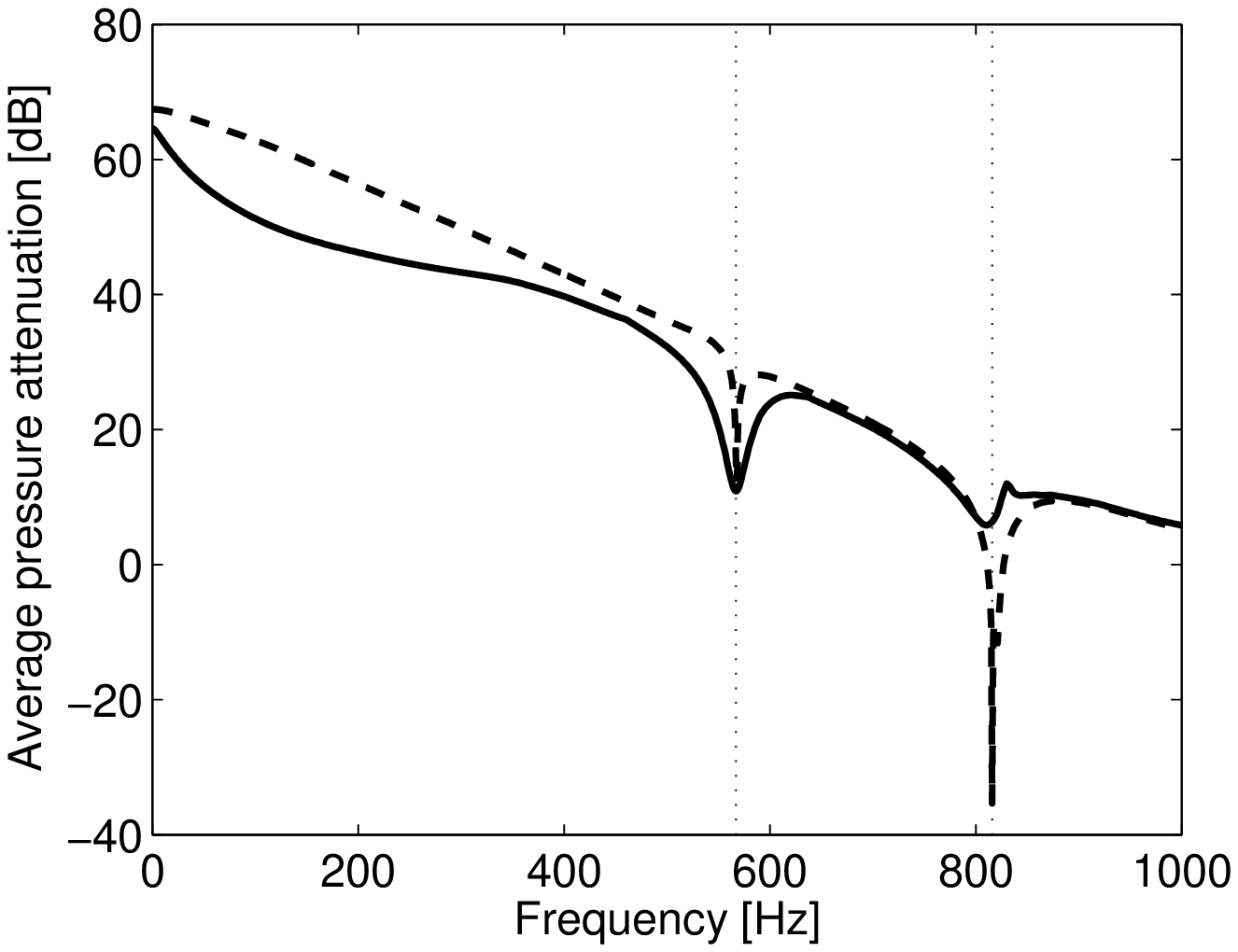}
  \vspace*{5cm}
  \caption{}
  \label{simul__freq__att_moy_volu_comp}
\end{figure}
\begin{figure}
  \vspace*{4cm} \centering
  \includegraphics[width=12cm]{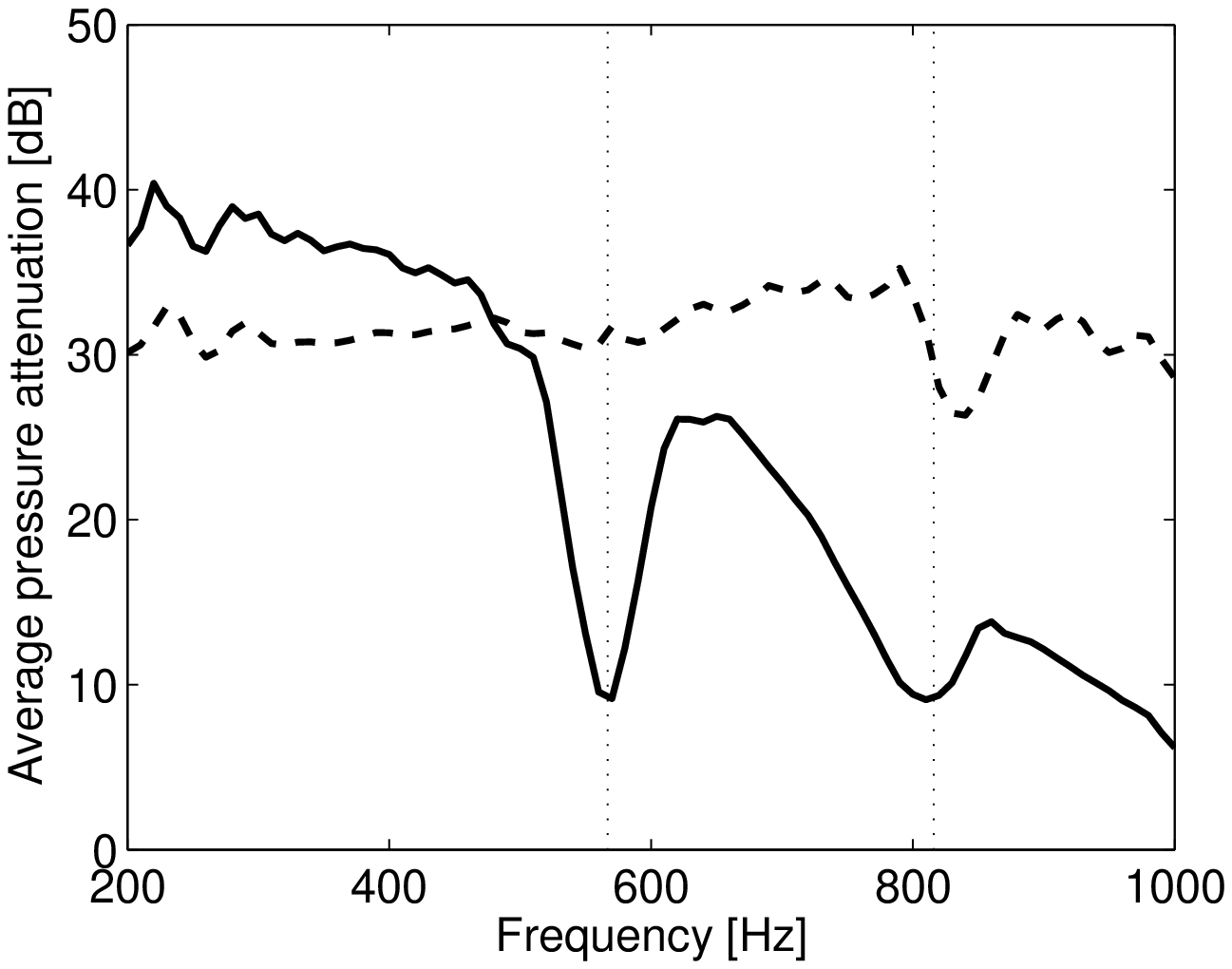}
  \vspace*{5cm}
  \caption{}
  \label{simul__temp__sweptsine_att_moy}
\end{figure}
\begin{figure}
  \vspace*{4cm} \centering
  \includegraphics[width=12cm]{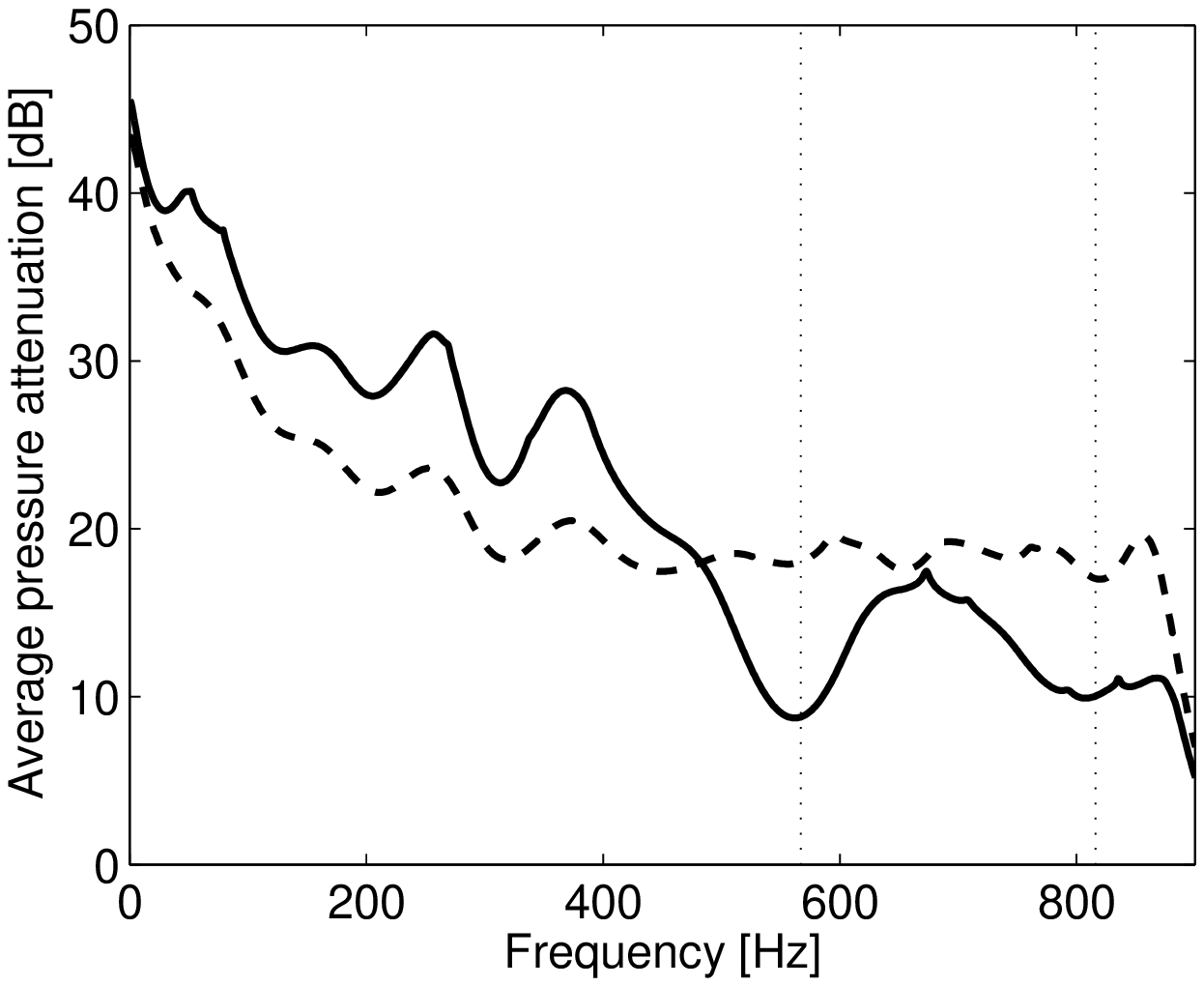}
  \vspace*{5cm}
  \caption{}
  \label{simul__temp__broadband}
\end{figure}
\begin{figure}
  \vspace*{4cm} \centering
  \includegraphics[width=12cm]{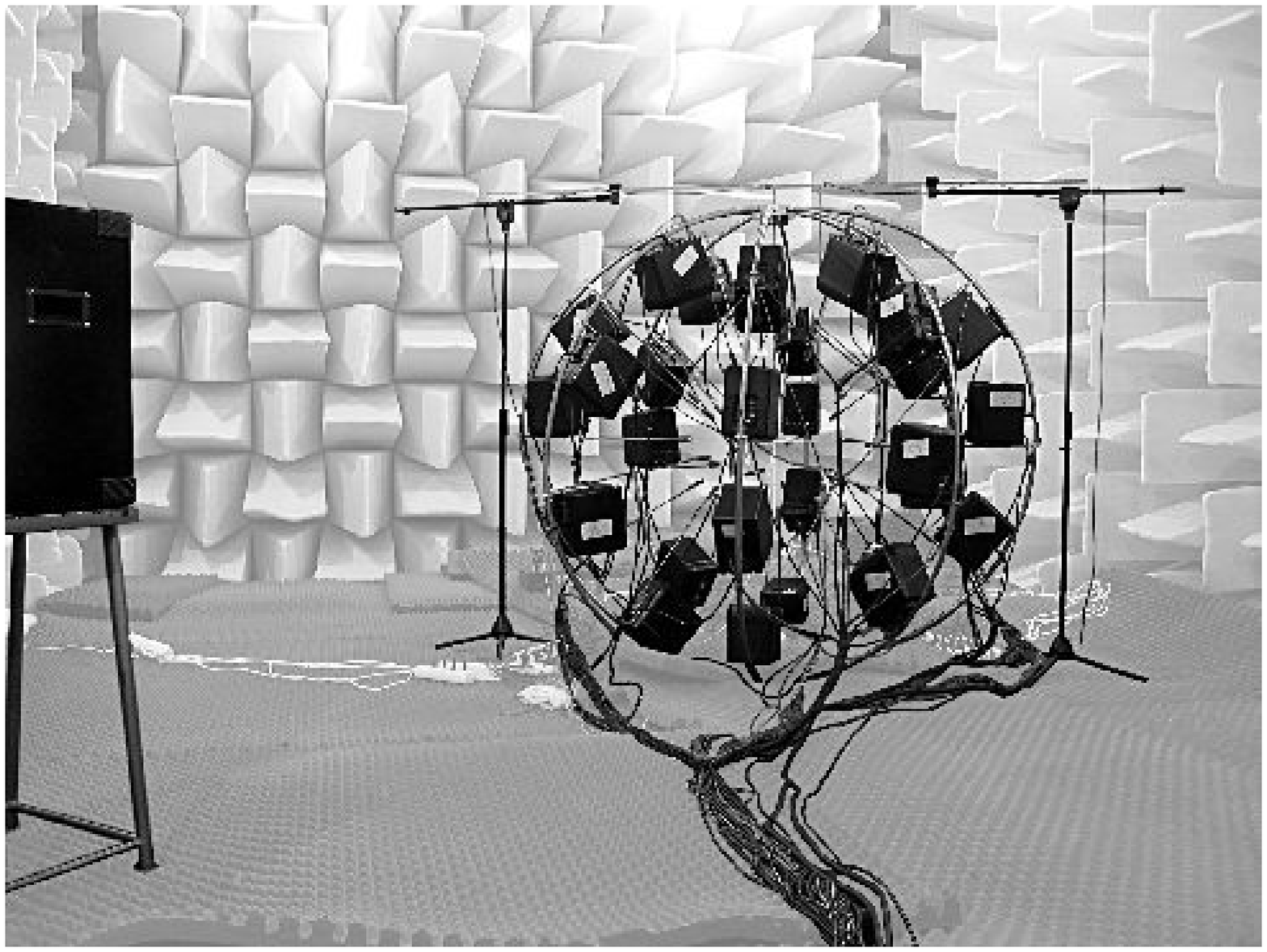}
  \vspace*{5cm}
  \caption{}
  \label{expe__photo_1}
\end{figure}
\begin{figure}
  \vspace*{4cm} \centering
  \includegraphics[width=12cm]{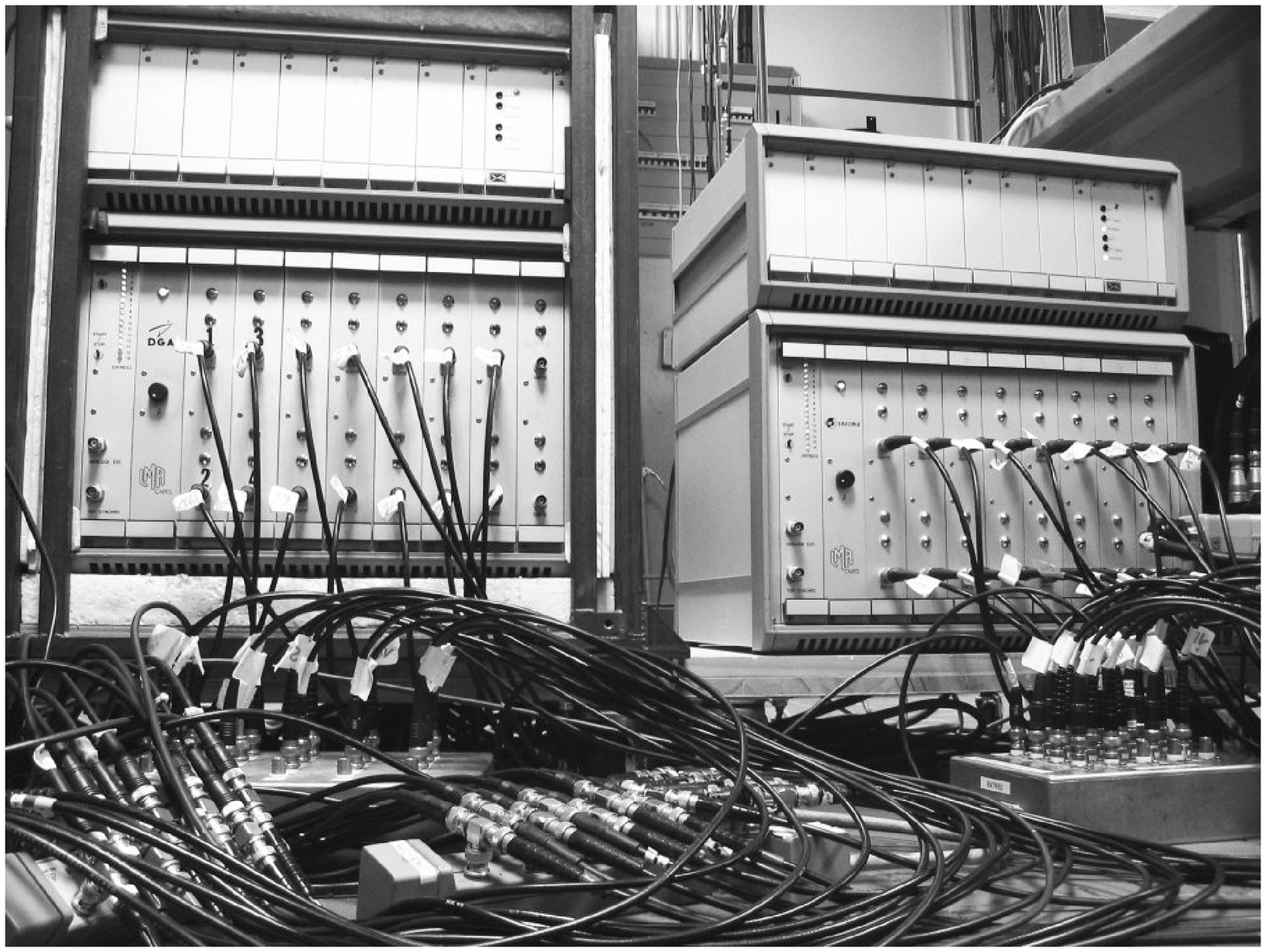}
  \vspace*{5cm}
  \caption{}
  \label{expe__photo_2}
\end{figure}
\begin{figure}
  \vspace*{4cm} \centering
  \includegraphics[width=12cm]{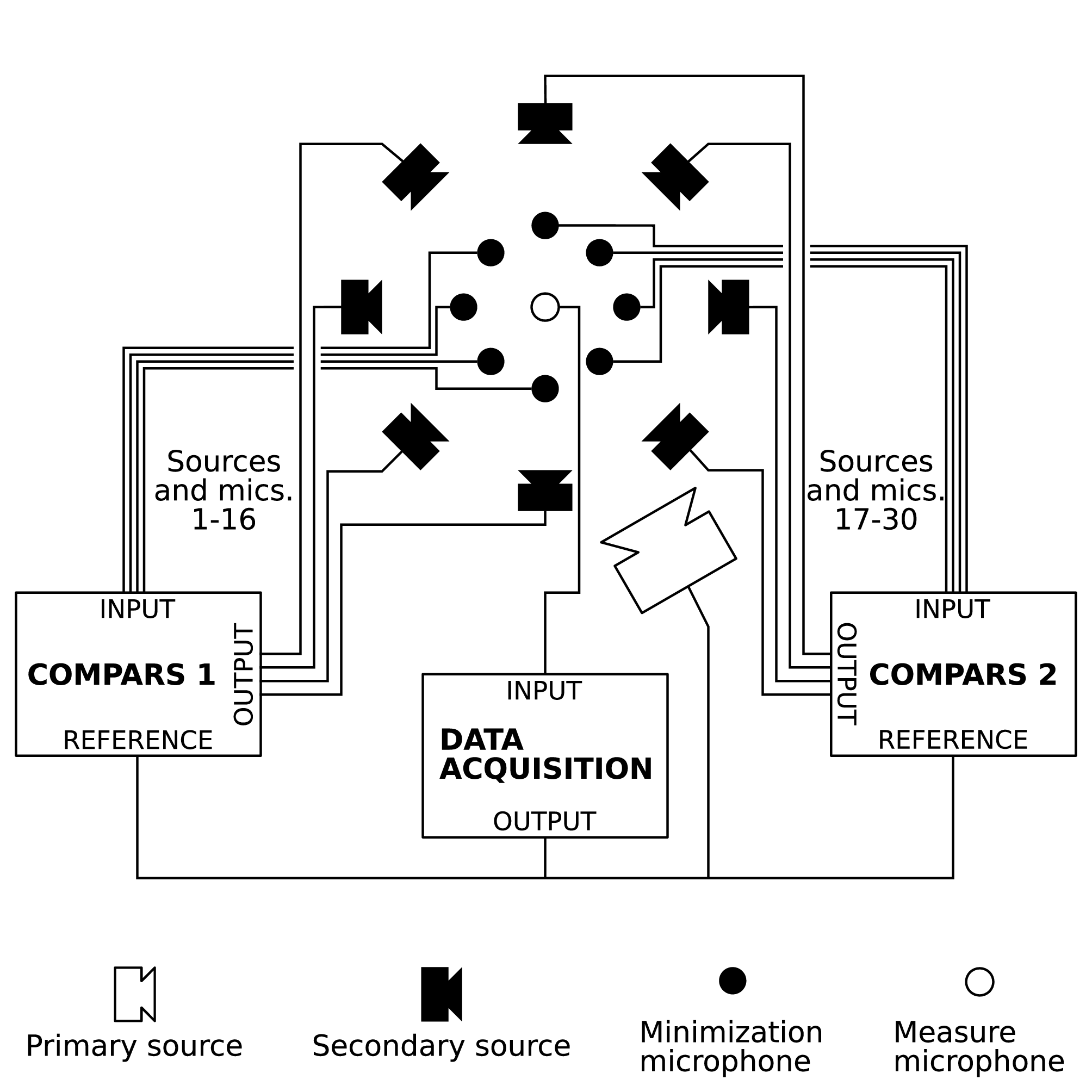}
  \vspace*{5cm}
  \caption{}
  \label{expe__broadband__setup}
\end{figure}
\begin{figure}
  \vspace*{4cm} \centering
  \includegraphics[width=12cm]{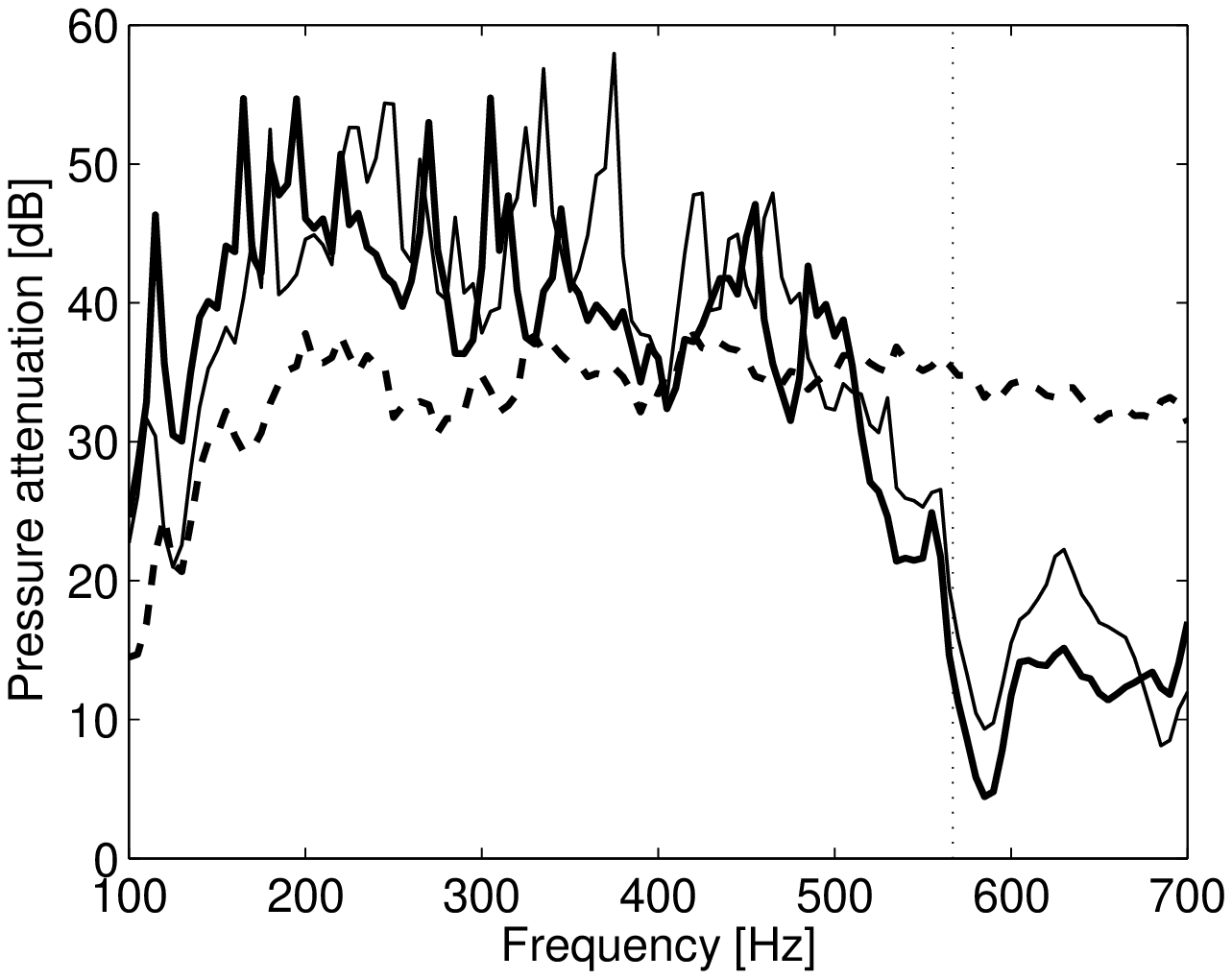}
  \vspace*{5cm}
  \caption{}
  \label{expe__sweptsine__att}
\end{figure}
\begin{figure}
  \vspace*{4cm} \centering
  \includegraphics[width=12cm]{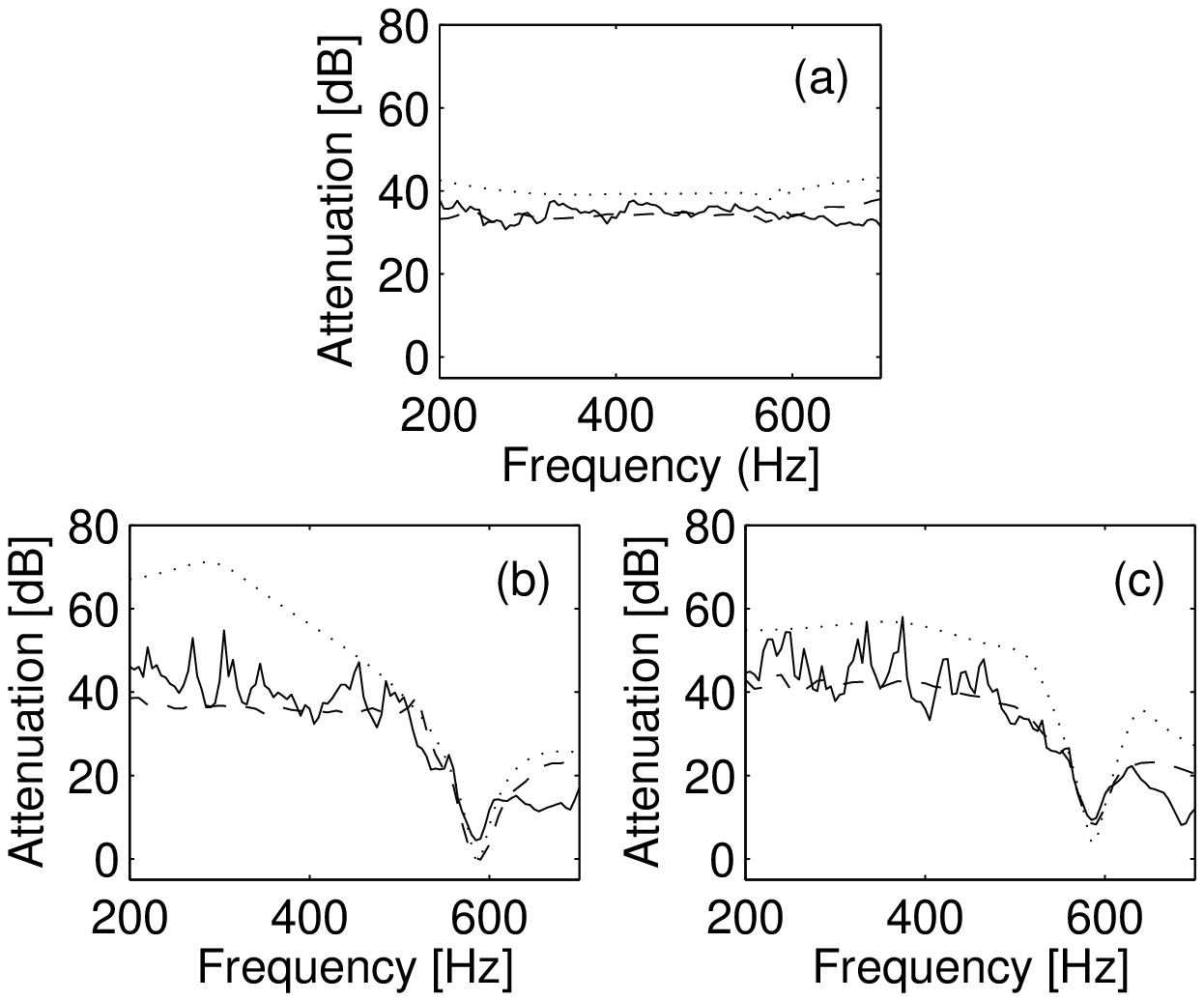}
  \vspace*{5cm}
  \caption{}
  \label{expe__sweptsine__comp}
\end{figure}
\begin{figure}
  \vspace*{4cm} \centering
  \includegraphics[width=12cm]{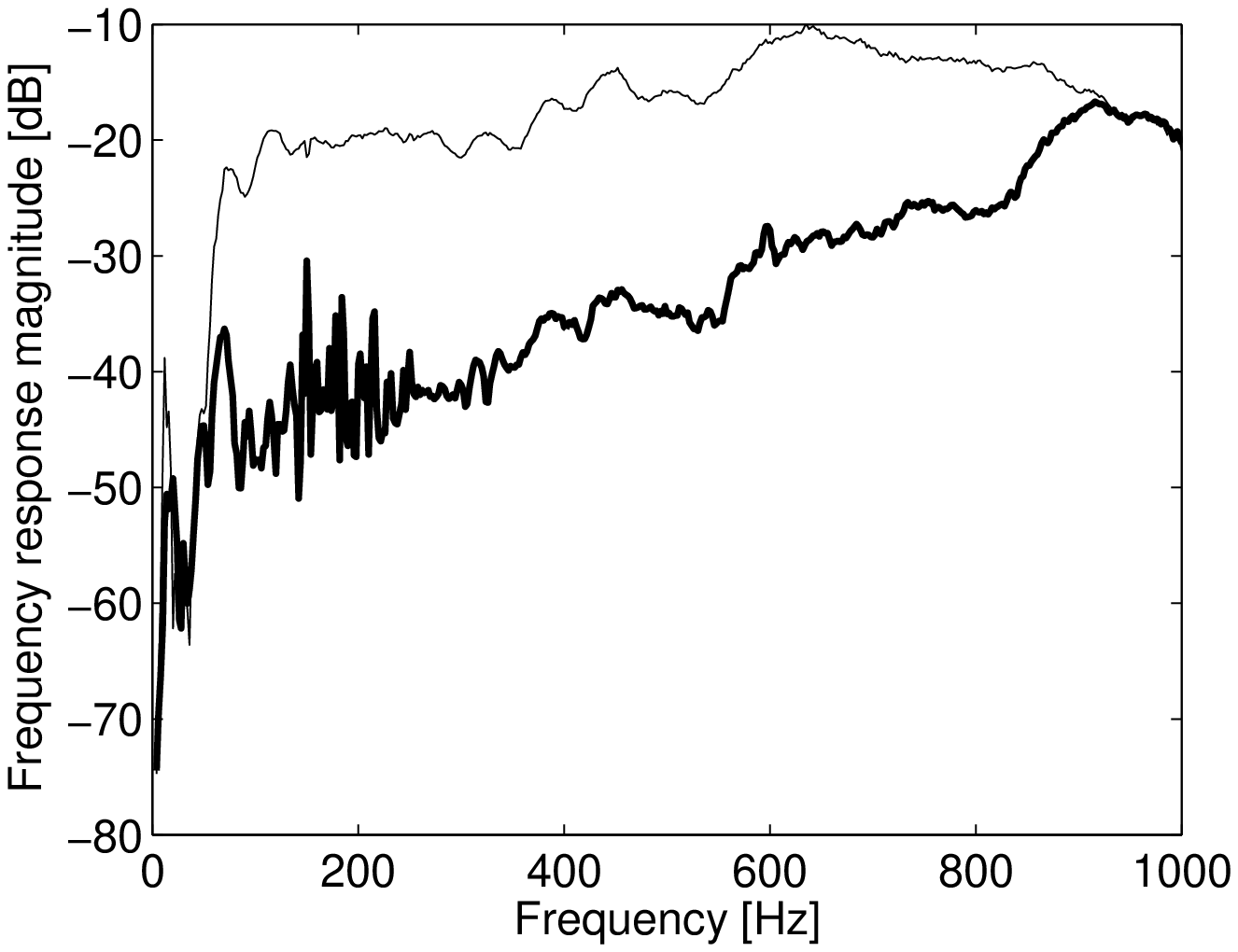}
  \vspace*{5cm}
  \caption{}
  \label{expe__broadband__frf}
\end{figure}
\begin{figure}
  \vspace*{4cm} \centering
  \includegraphics[width=12cm]{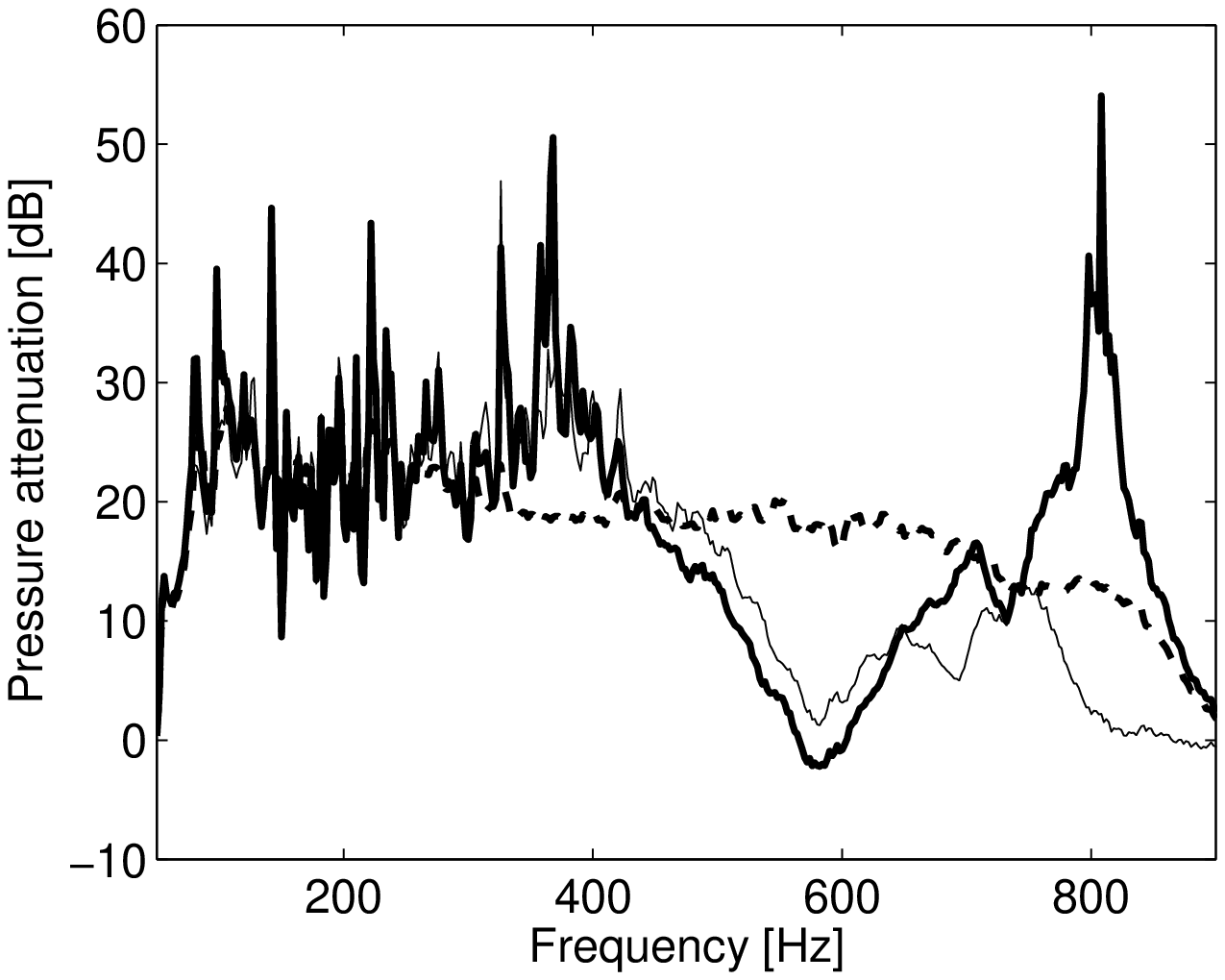}
  \vspace*{5cm}
  \caption{}
  \label{expe__broadband__att}
\end{figure}
\begin{figure}
  \vspace*{4cm} \centering
  \includegraphics[width=12cm]{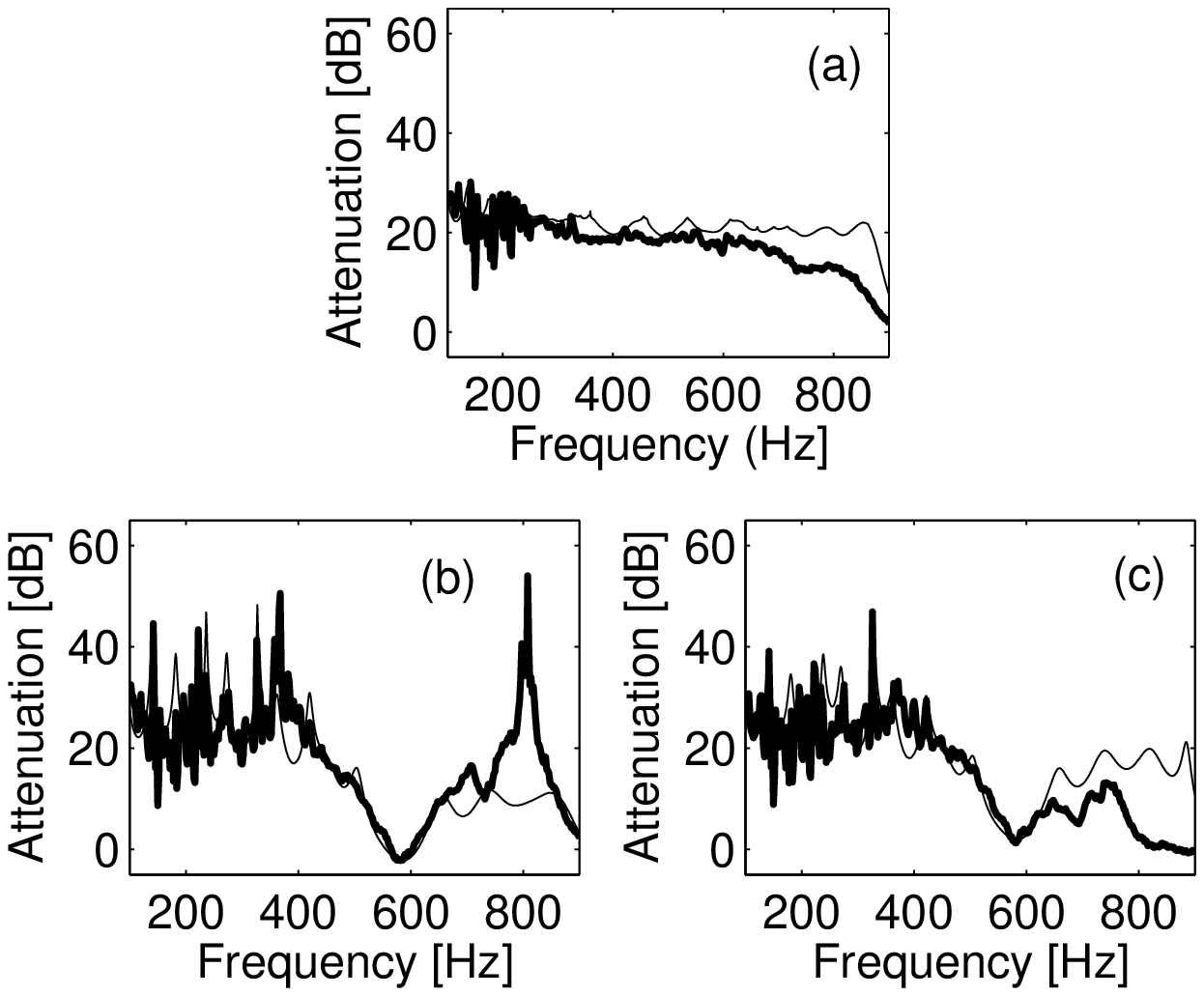}
  \vspace*{5cm}
  \caption{}
  \label{expe__broadband__comp}
\end{figure}

Fig. \ref{kirchfig}: Notations for the Kirchhoff--Helmholtz equation \\
Fig. \ref{simul__sphere_mic_fig}: The microphone arrangement of the simulation setup \\
Fig. \ref{simul__setup_fig}: Geometry and dimensions of the simulation setup \\
Fig. \ref{simul__freq__att_moy_volu}: Frequency-domain simulation: average pressure attenuation measured inside the sphere as a function of the frequency, without regularization. The dotted lines mark the two first eigenfrequencies of the interior Dirichlet problem, at $566$ and $816$~Hz. \\
Fig. \ref{simul__freq__att_moy_volu_reg}: Frequency-domain simulation: average pressure attenuation as a function of the frequency, with regularization (\rule[2pt]{12pt}{1pt} inside the sphere, \textbf{{-}{-}{-}} at the minimisation microphones)\\
Fig. \ref{simul__freq__att_moy_volu_comp}: Frequency-domain simulation: average pressure attenuation measured inside the sphere as a function of the frequency (\rule[2pt]{12pt}{1pt} with regularization,  \textbf{{-}{-}{-}} without regularization) \\
Fig. \ref{simul__temp__sweptsine_att_moy}: Time-domain simulation: average pressure attenuation obtained for pure tone signals as a function of the frequency (\rule[2pt]{12pt}{1pt} inside the sphere, \textbf{{-}{-}{-}} at the minimisation microphones) \\
Fig. \ref{simul__temp__broadband}: Time-domain simulation: average pressure attenuation obtained in the case of broadband noise as a function of the frequency (\rule[2pt]{12pt}{1pt} inside the sphere, \textbf{{-}{-}{-}} minimisation microphones)\\
Fig. \ref{expe__photo_1}: The experimental setup. On the left: the primary source. On the right: the 30 secondary sources, the 30 minimisation microphones, and the two interior measurement microphones.\\
Fig. \ref{expe__photo_2}: The two COMPARS controllers used in the case of the broadband noise signal.\\
Fig. \ref{expe__broadband__setup}: Experimental setup used in the case of the broadband noise signal.\\
Fig. \ref{expe__sweptsine__att}: Experimental results: the pressure attenuation measured in the case of for pure-tone primary signals as a function of the frequency (\rule[2pt]{12pt}{1pt} interior microphone 1, --- interior microphone 2, \textbf{{-}{-}{-}} minimisation microphones) \\
Fig. \ref{expe__sweptsine__comp}: Comparison between simulation and experimental results, in the case of pure-tone primary signals: (a) average attenuation at the minimisation microphones, (b) interior microphone 1, (c) interior microphone 2 (--- experiment, {-}{-}{-} time-domain simulation, $\cdots$ frequency-domain simulation)\\
Fig. \ref{expe__broadband__frf}: Experimental results: magnitude of the frequency response measured between the primary source command and the minimisation microphones in the case of a broadband noise primary signal (\rule[2pt]{12pt}{1pt} control on, --- control off)\\
Fig. \ref{expe__broadband__att}: Experimental results: pressure attenuation measured with a broadband noise primary signal as a function of the frequency (\rule[2pt]{12pt}{1pt} interior microphone 1, --- interior microphone 2, \textbf{{-}{-}{-}} minimisation microphones)\\
Fig. \ref{expe__broadband__comp}: Comparison between simulation and
experimental results, in the case of a broadband noise primary signal:
(a) average attenuation at the minimisation microphones, (b) interior
microphone 1, (c) interior microphone 2 (\rule[2pt]{12pt}{1pt}
experiment, --- simulation)

\newpage

\newpage

\newpage

\newpage

\newpage

\newpage

\newpage

\newpage

\newpage

\newpage

\newpage

\newpage

\newpage

\newpage

\newpage

\newpage

\end{document}